\documentclass[12pt,twoside,frenchb,english]{article}
\usepackage{bookman}
\usepackage[T1]{fontenc}
\usepackage[latin1]{inputenc}
\usepackage{amsmath}
\usepackage{graphicx}
\usepackage{amssymb}

\makeatletter

\providecommand{\LyX}{L\kern-.1667em\lower.25em\hbox{Y}\kern-.125emX\@}

\usepackage{bookman}
\usepackage{a4}
\usepackage{cite}
\input{epsfig.sty}
\def\fnum@table{\tablename~{\bf\thetable}}
\def\fnum@figure{\figurename~{\bf\thefigure}}
\def\tablename{\footnotesize{\bf Table}}
\def\figurename{\footnotesize{\bf Figure}}

\AtBeginDocument{

}

\usepackage{babel}
\makeatother
\begin{document}
\vspace{-0.5cm}
\begin{center}\textbf{\huge Particle Production in Proton-Proton and
Deuteron-Gold Collisions at RHIC}\end{center}{\huge \par}

\begin{center}{\Large Klaus Werner$^{(a)}$}%
\footnote{Invited talk, given at \foreignlanguage{frenchb}{SQM2004, Cape Town,
South Africa, 15-20 September, 2004}%
}{\Large , Fuming Liu$^{(b)}$ , Tanguy Pierog$^{(c)}$}\end{center}{\Large \par}

{\footnotesize $^{(a)}$ SUBATECH, Université de Nantes -- IN2P3/CNRS
-- Ecole des Mines, Nantes, France }\\
{\footnotesize $^{(b)}$ Institute of Particle Physics , Huazhong
Normal University, Wuhan, China}\\
 {\footnotesize $^{(c)}$ FZK, Institut für Kernphysik, Karlsruhe,
Germany}{\footnotesize \par}

\begin{abstract}
We try to understand recent data on proton-proton and deuteron-gold
collisions at RHIC, employing a parton model approach called EPOS.
\end{abstract}

\section{Introduction}

EPOS is a new energy conserving multiple scattering approach based
on partons and Pomerons (parton ladders), with special emphasis on
high \underbar{}parton \underbar{}densities. \underbar{}The latter
aspect, particularly important in proton-nucleus or nucleus-nucleus
collisions, is taken care of via an effective treatment of Pomeron-Pomeron
interactions. Soft and hard interactions are treated in a consistent
way. EPOS is the successor of the NEXUS model.

\section{High parton densities}

\begin{figure}[htb]
\begin{center}\begin{minipage}[c]{0.10\textwidth}%
(A) \\
\\
\\
\\
\end{minipage}%
\begin{minipage}[c]{0.10\textwidth}%
\includegraphics[  scale=0.7]{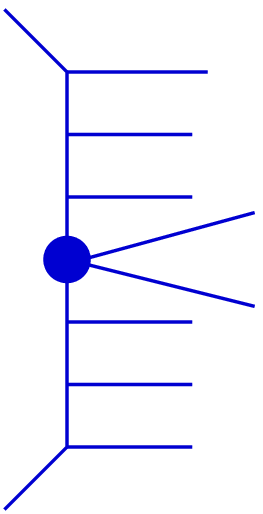}\end{minipage}%
\begin{minipage}[c]{0.10\textwidth}%
$\quad \to \quad $\end{minipage}%
\begin{minipage}[c]{0.10\textwidth}%
\includegraphics[  scale=0.7]{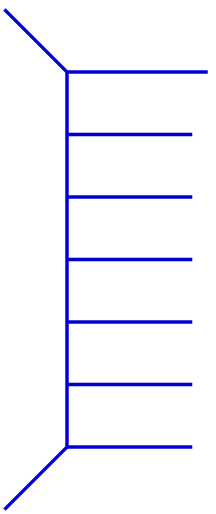}\end{minipage}%
$\qquad \qquad \qquad $\begin{minipage}[c]{0.10\textwidth}%
(B) \\
\\
\\
\\
\end{minipage}%
\begin{minipage}[c]{0.30\textwidth}%
\includegraphics[  scale=0.7]{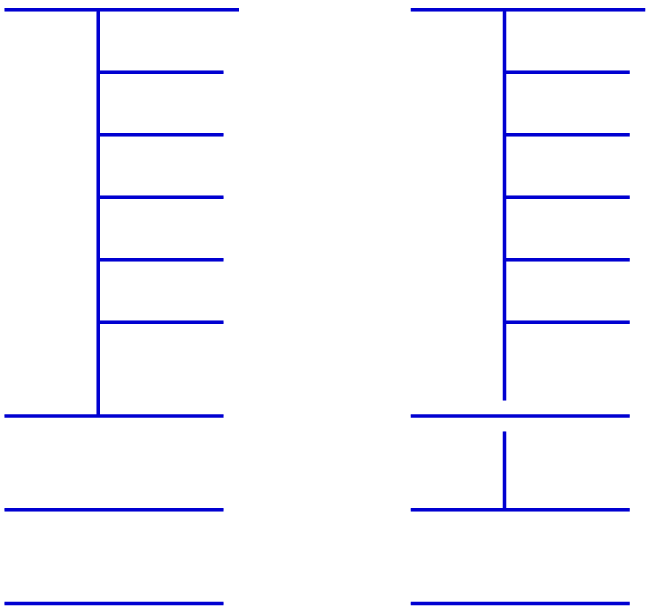}\end{minipage}%
\end{center}

\caption{(A) Parton-parton scattering. \label{cap:Parton-parton}~(B) Scattering
with many partons. }
\end{figure}

Let us first consider parton-parton scattering. A parton from the
projectile, after emitting several (initial state) partons, interacts
with a corresponding parton from the target, see figure \ref{cap:Parton-parton}A
(left). To simplify graphical representations in the following, we
use a symbolic parton ladder , as shown in figure \ref{cap:Parton-parton}A
(right), representing both soft and hard interactions.

Having several target partons available, the projectile parton may
interact in this way with any of the target partons, as shown in fig.
\ref{cap:Parton-parton}B, which will simply change the cross section
by some factor.

\begin{figure}[htb]
\begin{center}\begin{minipage}[c]{0.10\textwidth}%
(A) \\
\\
\\
\\
\\
\\
\end{minipage}%
\begin{minipage}[c]{0.10\textwidth}%
\includegraphics[  scale=0.7]{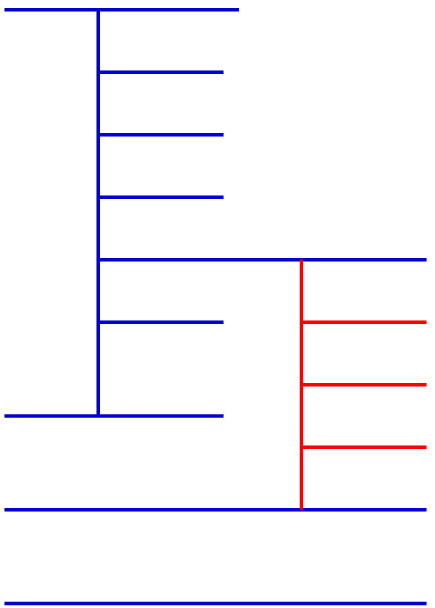}\end{minipage}%
$\qquad \qquad \qquad \qquad $\begin{minipage}[c]{0.10\textwidth}%
(B) \\
\\
\\
\\
\\
\\
\end{minipage}%
\begin{minipage}[c]{0.20\textwidth}%
\includegraphics[  scale=0.7]{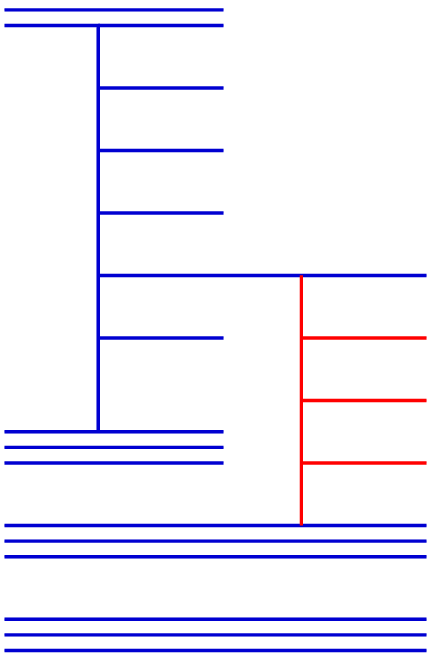}\end{minipage}%
$\quad \to \quad $\begin{minipage}[c]{0.15\textwidth}%
\includegraphics[  scale=0.7]{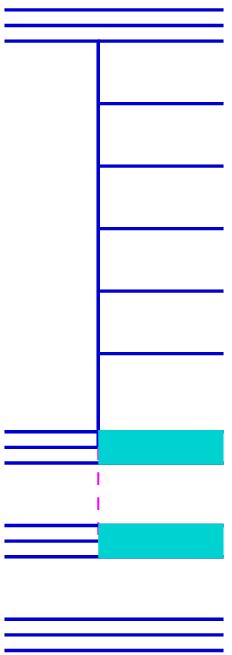}\end{minipage}%
\end{center}

\caption{(A) Multiple ladders.\label{cap:hiden}~ (B) Effective treatment
of multiple ladder contributions.}
\end{figure}

The situation will, however, be more complicated in case of high parton
densities. Here, a parton from a ladder may rescatter with another
parton from the projectile or target, providing an additional ladder
(fig. \ref{cap:hiden}A). %
 ~
\vspace{2cm}

We try to model such high density effects

\begin{itemize}
\item in a simple and transparent way,
\item using just simple ladders between projectile and target (Pomerons),
\item putting all complications into {}``projectile / target excitations'',
to be treated in an effective way,
\end{itemize}
as shown in fig.\ref{cap:hiden}B. Bifurcation of parton ladders will
not be treated explicitly, they are absorbed into target and projectile
excitations, visualized as fat lines in the figure. The excitations
may represent one, two, or even more ladders, depending on the parton
densities.

\textbf{How to realize these ideas?}

For a given target nucleon $j_{0}$, we define the number $Z_{\mathrm{T}}(j_{0})$
of additional partons on the target side being available for multiple
interactions, as seen by the projectile nucleons interacting with
$j_{0}$ :\[
Z_{\mathrm{T}}(j_{0})=\sum _{\begin{array}{c}
 \mathrm{proj.nucleons}\, i\\
 \mathrm{interacting}\\
 \mathrm{with}\, j_{0}\end{array}
}\max \big (0,\sum _{\begin{array}{c}
 \mathrm{target}\\
 \mathrm{nucleons}\, j\end{array}
}z(b_{ij})\; -1\big )\]
with\[
z(b)=\frac{z_{0}(E)}{\sqrt{b_{\mathrm{cut}}^{2}+b^{2}}}\Theta \big (\sqrt{\frac{\sigma _{\mathrm{inel}}}{\pi }}-b\big ).\]
A corresponding definition holds for $Z_{P}$. For pA: $Z_{\mathrm{P}}\approx 0$,
$Z_{\mathrm{T}}\propto N_{\mathrm{target}\, \mathrm{participants}}$.

\textbf{How to realize projectile / target excitations ?} (accounting
for multiple, interacting ladders) 

We suppose an excitation mass distributed according to $1/M^{2(1-\mu )}$.
Low masses correspond to hadrons or resonances, high mass excitations
are considered to be strings.

The string parameters $\left\langle \right.p_{t}^{\mathrm{string}\, \mathrm{break}}\left.\right\rangle $and
$p_{\mathrm{strange}}$, as well as $\mu $ and $\left\langle \right.p_{t}^{\mathrm{string}\, \mathrm{end}}\left.\right\rangle $
of the strings connected to an excited target/projectile, depend on
$Z$ as $f=f_{0}\, \min (Z_{\max },1+aZ)$.

To provide some numbers: $Z_{\mathrm{T}}(\mathrm{centr}\, \mathrm{dAu})\approx 3-4,\, Z_{\mathrm{T}}(\mathrm{periph}\, \mathrm{dAu})\approx 0$,
$Z_{\max }\approx 4$, $a\approx 1$.

The formalism is based on cut diagram techniques, strict energy conservation,
and Markov chains for the numerics \cite{nexus}.

\section{Some results}

\begin{figure}[htb]
\begin{center}\includegraphics[  scale=0.25,
  angle=270,
  origin=c]{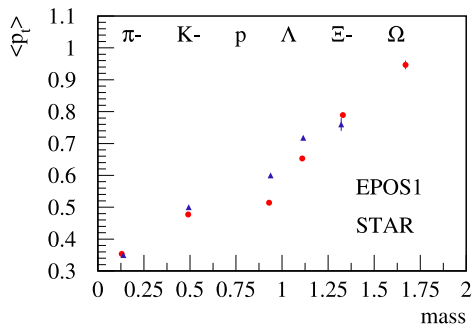}\end{center}
\vspace{-1.7cm}

\caption{Mean transverse momenta in pp collisions; data (triangles) from ref.
\cite{star}.\label{cap:Mean-transverse-momenta}}
\end{figure}

Very detailed tests show that the model works very well for pp scattering.
As an example, we show in fig. \ref{cap:Mean-transverse-momenta}
mean transverse momenta for different hadron species.%

\begin{figure}[htb]
\begin{center}\includegraphics[  scale=0.23,
  angle=270,
  origin=c]{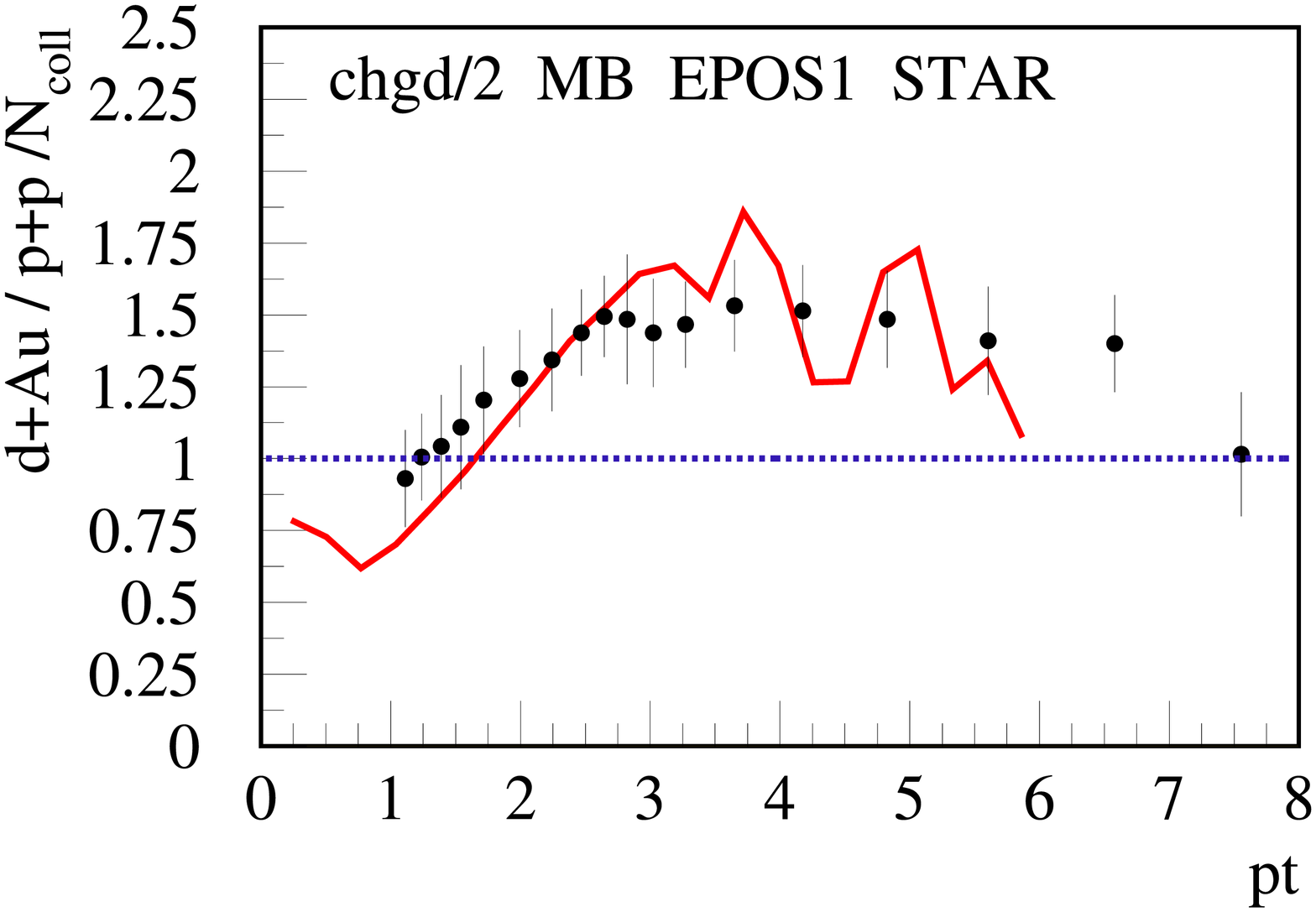}\includegraphics[  scale=0.23,
  angle=270,
  origin=c]{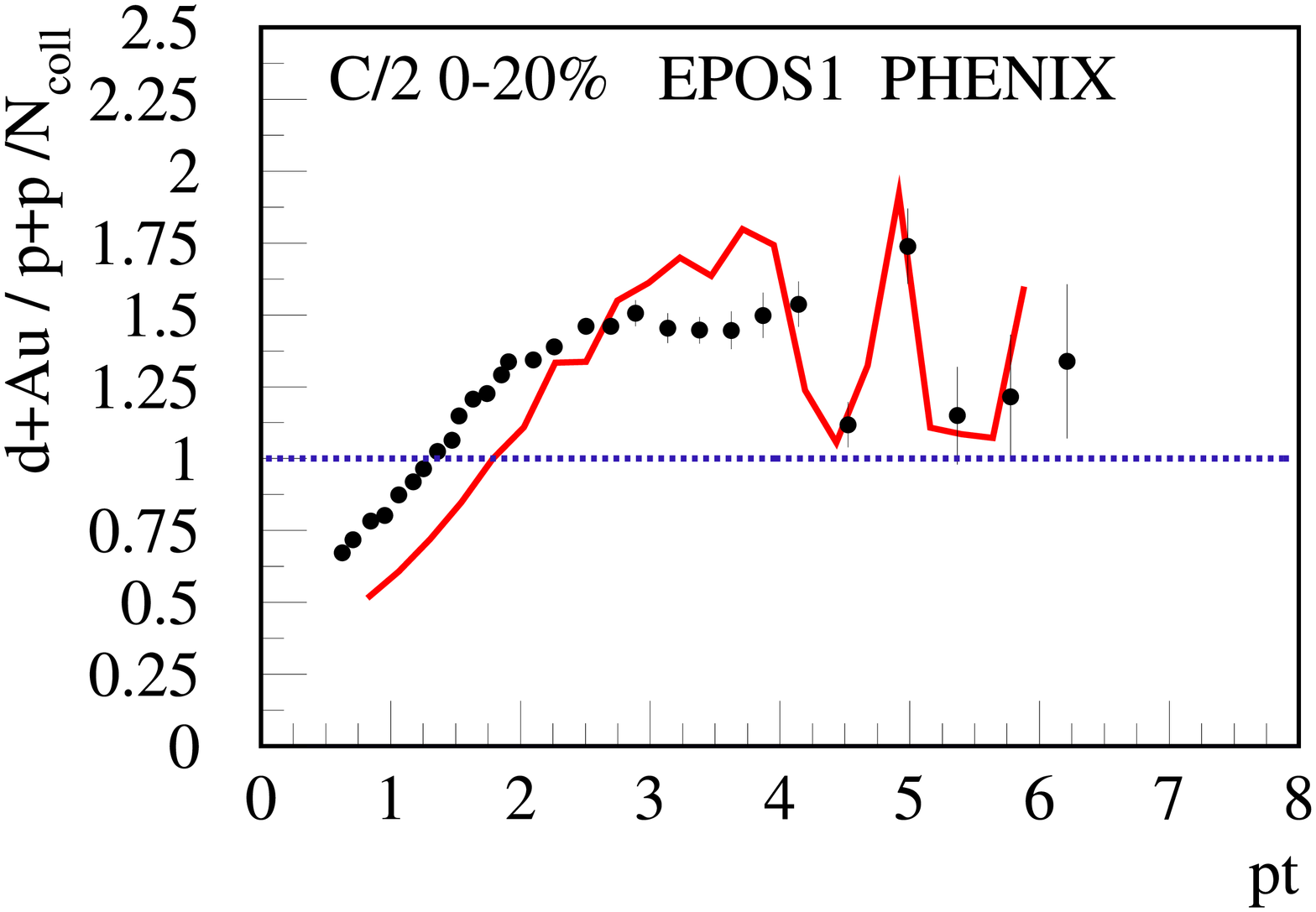}\end{center}
\vspace{-1.7cm}

\caption{The nuclear modification factor; data (points) from ref.\cite{star,phenix}.
\label{cap:Nuclear-modification-factor}}
\end{figure}

Concerning the nuclear modification factor $R=AA/pp/N_{\mathrm{coll}}$
as a function of $p_{t}$, in d+Au collisions, we observe an increase
beyond one, as observed in the data, see fig. \ref{cap:Nuclear-modification-factor},
due to the fact that $Z_{T}$ becomes quite big (up to 3-4).%

\begin{figure}[htb]
\begin{center}\includegraphics[  scale=0.23,
  angle=270,
  origin=c]{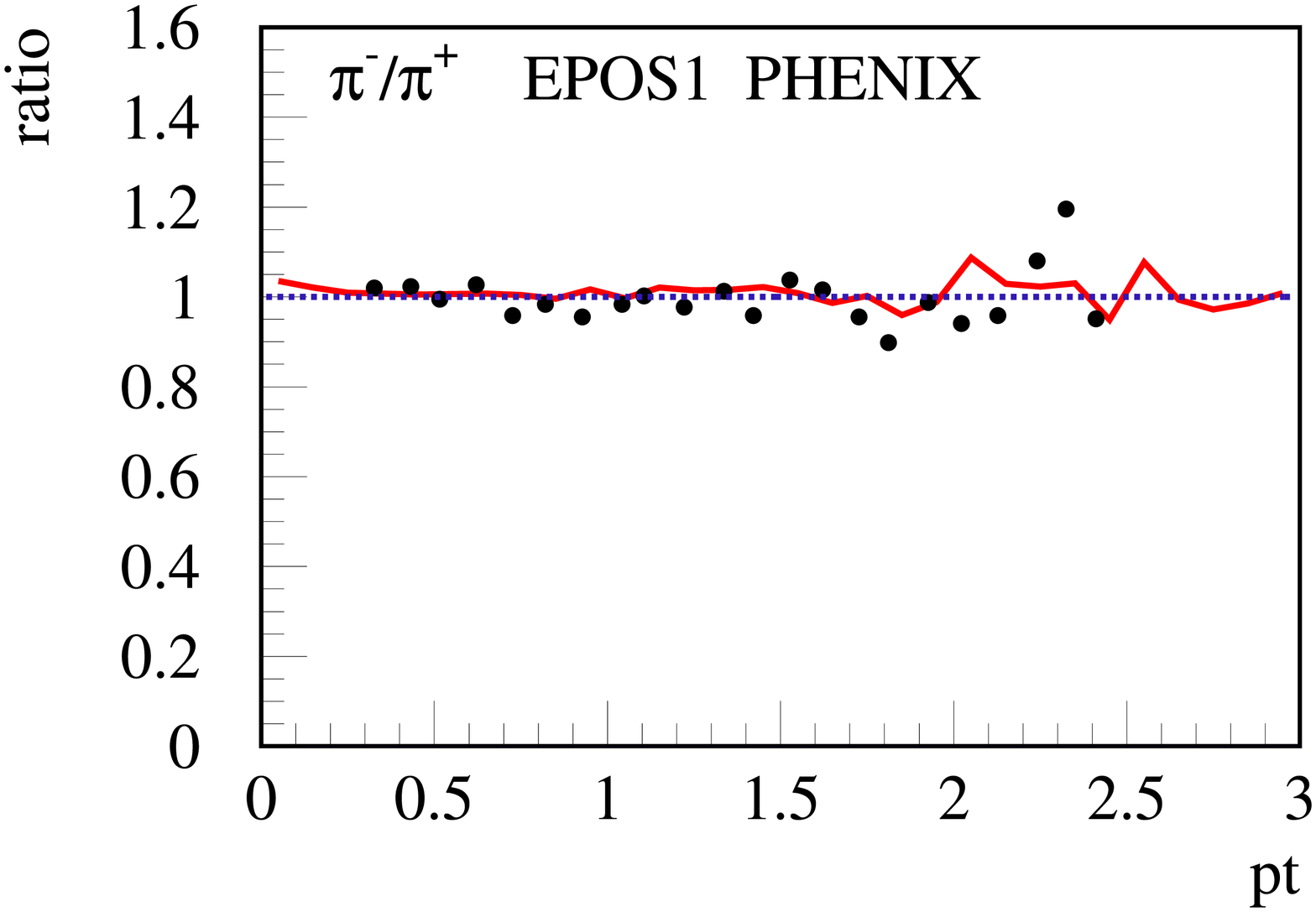}\includegraphics[  scale=0.23,
  angle=270,
  origin=c]{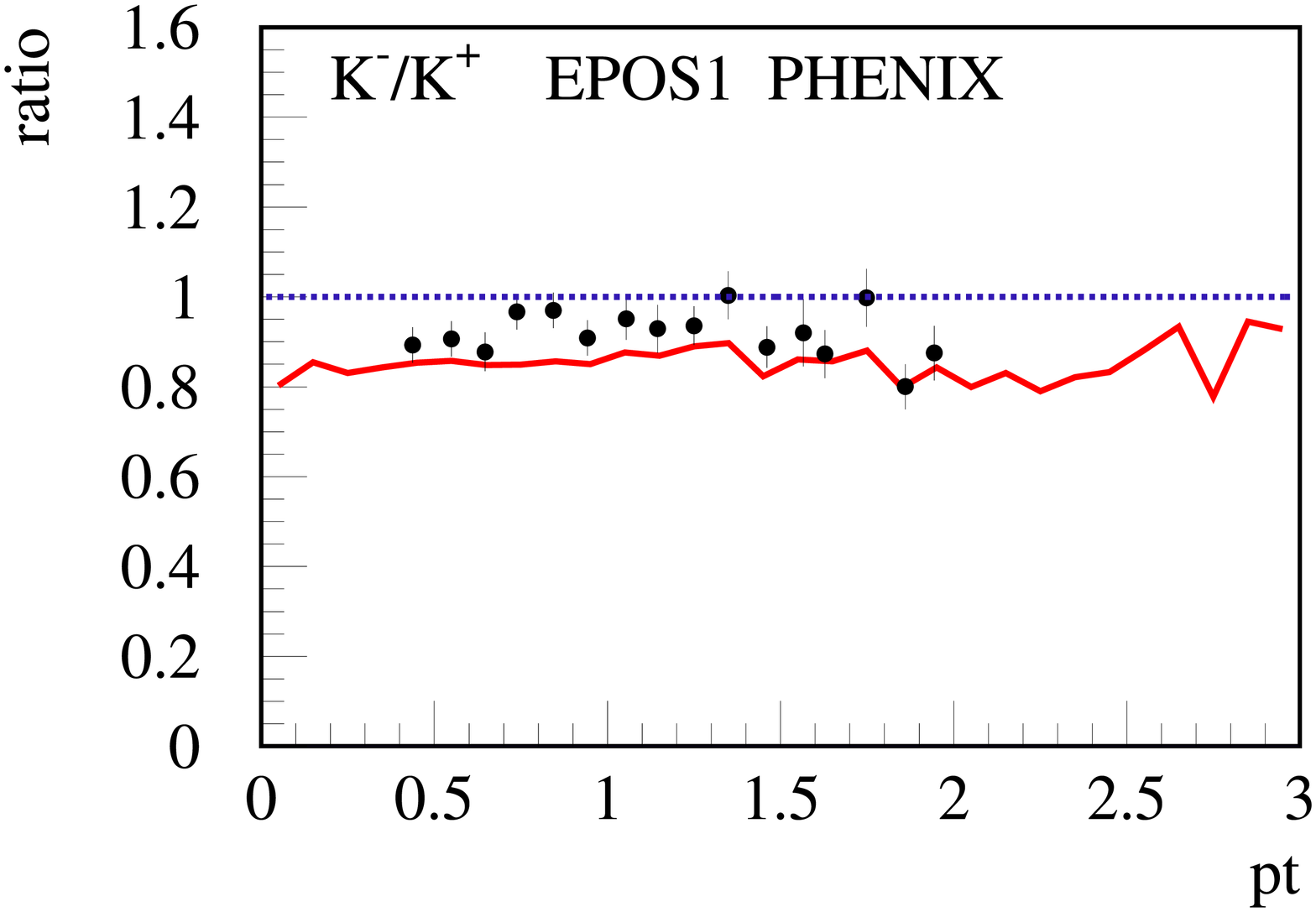}\end{center}
\vspace{-1.8cm}

\begin{center}\includegraphics[  scale=0.23,
  angle=270,
  origin=c]{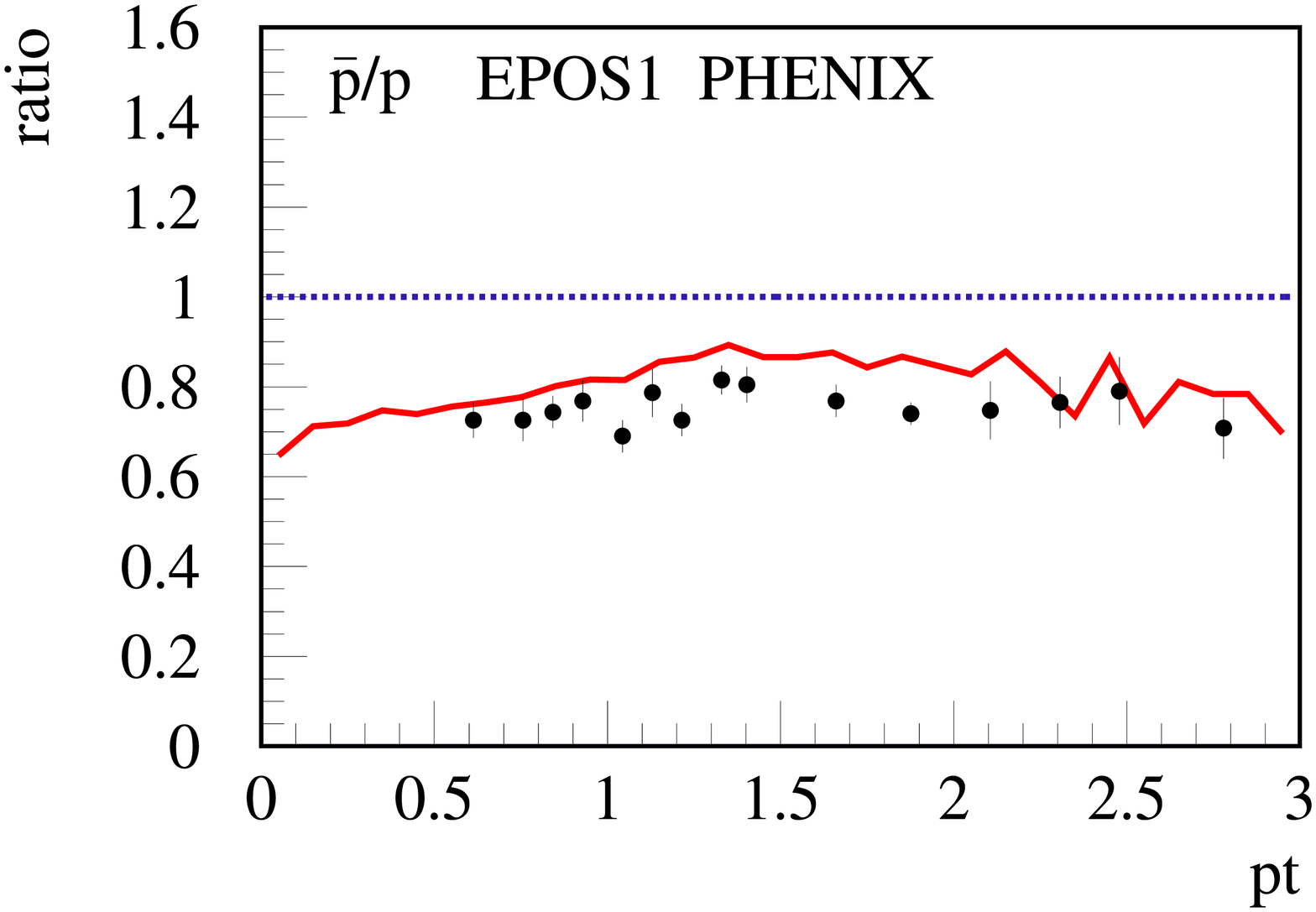}\end{center}
\vspace{-1.8cm}

\caption{Transverse momentum dependence of hadron ratios; data from ref. \cite{phenix}.\label{cap:Transverse-momentum-dependence}}
\end{figure}

In fig. \ref{cap:Transverse-momentum-dependence} finally, we show
the transverse momentum dependence of hadron ratios. The pt dependence
in mainly determined by the mass of the particle, so ratios of particles
of equal mass are essentially constant.


\begin{thebibliography}{1}
\bibitem{nexus}H.J. Drescher, M. Hladik, S. Ostapchenko, T. Pierog, K. Werner, Phys.
Rept. 350 (2001) 93
\bibitem{star}STAR, Phys. Rev. Lett. 91 (2003) 72304, and contributions to this
conference
\bibitem{phenix}PHENIX collaboration, Phys. Rev. Lett. 91:072303 (2003), and F. Matathias,
presentation at Quark Matter 2004. \end{thebibliography}
\end{document}